# Breviz: Visualizing Spreadsheets using Dataflow Diagrams

Felienne Hermans, Martin Pinzger, Arie van Deursen
Delft University of Technology, Mekelweg 4, Delft, The Netherlands
{f.f.j.hermans,m.pinzger,arie.vandeursen}@tudelft.nl

## ABSTRACT

*Spreadsheets are used extensively in industry, often for business critical purposes. In previous work we have analyzed the information needs of spreadsheet professionals and addressed their need for support with the transition of a spreadsheet to a colleague with the generation of data flow diagrams. In this paper we describe the application of these data flow diagrams for the purpose of understanding a spreadsheet with three example cases. We furthermore suggest an additional application of the data flow diagrams: the assessment of the quality of the spreadsheet's design.*

## 1. INTRODUCTION

Spreadsheets are used widely in industry, for all kinds of tasks, like financial modeling, reporting and planning [Rittweger, 2010].A study from the year 2005 shows about 23 million American workers use spreadsheets, which amounts toabout 30% of the workforce[Scaffidi, 2005].

Many of the spreadsheet used are of great importance to companies, Hall [Hall, 1996] interviewed 106 spreadsheet developers and found that only 7% of the spreadsheets were of low importance and that as much as 39% were of high importance. In a recent study we found similar results [Hermans, 2011].

In that study we furthermore investigated the most prevalent problems spreadsheet users have in their daily work with spreadsheets, by interviewing 27 spreadsheet users working at Robeco, a Dutch investment bank. The results showed that problems arose when spreadsheets are transferred from one employee to another. As spreadsheets have an average lifetime of more than five years, and individual spreadsheets are used by 13 different employees, this happens quite frequently. We identified three different transfer scenarios: from one employee to another, from an employee to IT, and from an employee to an auditor [Hermans, 2011].

In the case of a transfer, the receiving party often has to spend hours browsing through the spreadsheet to understand its structure and purpose, since only one third of the spreadsheets contains documentation. The transferring party often feels the spreadsheet is not very complicated, and does not spend enough time to explain it.We have addressed this problem by creating a data flow diagram visualization that can be used during a transfer scenario to support the explanation of the spreadsheet.

When the tool was finished, it was installed at Robeco, so employees could use it when needed. When we analyzed the use of Breviz at Robeco, we found that it was not only useful in the transfer scenarios, but also for individual comprehension: when a spreadsheet user analyzes a spreadsheet by themselves. This scenario occurs when it is not possible to ask the creator of the spreadsheet for advice, for instance when he left the company, or is on a holiday.

In this paper we briefly describe our data flow diagram generation approach, and explain how it can be useful for the individual comprehension of a spreadsheet, with three examples, based on our experiences in practice. We also discuss additional uses of our approach, for the identification of errors in the spreadsheet.

The remainder of this paper is structured as follows: Section 2 briefly explains the algorithm for the extraction of data flow diagram from a spreadsheet. Section 3 describes the three different views we support on those diagrams. In Section 4 the implementation of our approach into a tool---Breviz—is described. Section 5 subsequently explains the applicability of our approach with three practical examples. Section 6 discusses new applications of Breviz. Concluding remarks finally can be found in Section 7.

## 2. DATA FLOW DIAGRAMS CREATION

This section briefly explains how we extract a data flow diagram from a spreadsheet. For a more detailed description we refer to previous work [Hermans, 2011]. Dataflow diagrams---or similar techniques forrepresenting the flow within systems, such as flowcharts---havebeen present in literature since the seventies [Gane, 1977].

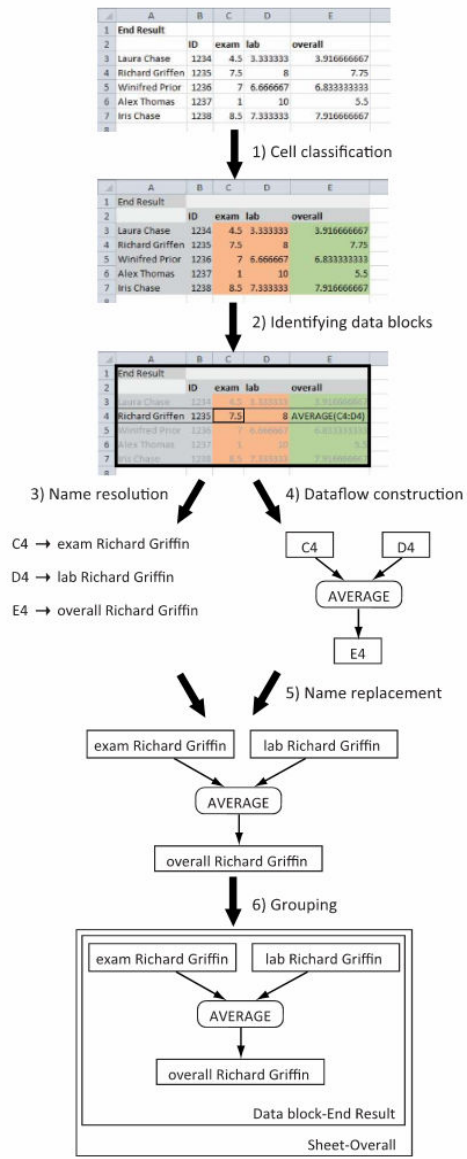

Figure 1 Overview of the data flow diagram extraction algorithm

The extraction of a data flow diagram from a spreadsheet is done in six steps, as illustrated by Figure 1.

The first two steps originate from our earlier work extracting class diagrams from spreadsheets[Hermans, 2010]. The first step determines the cell type of all cells in the spreadsheet, which can be Data, Formula, Label or Empty. The second step identifies data blocks within a spreadsheet. Data blocks are rectangles of non-empty cells in a worksheet, separated from other data blocks by empty cells.

In the third step, labels describing Data and Formulacells are computed. This is done by inspecting the borders of the data block the cell lies in. The fourth step generates an initial dataflow diagramby creating entities for cells of type Data andFormulaand creating arrows between them, corresponding to formula dependencies. Subsequently, in the fifth step,the labels of cells that were computed in step 3 are attached to the corresponding entities in the diagram The final step adds the levels to the dataflow diagram.A level is introduced for each worksheet within the spreadsheetand for each data block within every worksheet. Again we refer to our previous paper for the specifics of the data flow diagram generation.

## 3. DATA FLOW VIEWS

We support three different views on to the data flow diagram, to help users in navigating them.Examples of these views will be provided in Section 5.

The first view we support is the *global view*. This viewshows all worksheets within the spreadsheet and the relations between them. An arrow from worksheet A to worksheet B indicates a formula in worksheet B refers to a cell in worksheet A. Multiple arrows are grouped into one, so the thicker the arrow is, the more formulas reference cells of another worksheetThe second view is the *worksheet view*, which shows all data blocks within a worksheet, as well as thedependencies between them. The view is obtained from theglobal view by expanding a level representing a worksheet. In this way, details of a worksheetare revealed while keeping the overall picture of the spreadsheet. Finally there is the *formula view*, where the relation between formulas and the cells they depend on is shown.To obtain this view the user opens a data block-node in the worksheet view, showing all calculationsin the data block.

## 4. BREVIZ

Our initial prototype *GyroSAT* created the data flow diagram by generating a DGML file. The DGML (Directed Graph Markup Language) file format isan XMLschema for hierarchical directed graphs that can be viewed withthe graph browser that is part of Microsoft Visual Studio 2010 Ultimate. It is intended to visualize software architectures anddependency graphs for systems.

Because we did not have to implement a graph viewer ourselves, the use of DGML enabled us to create a prototype quickly. It however led to problems when we wanted to use the data flow diagram generation in an industrial setting, since not all spreadsheet users have Visual Studio installed, and were thus unable to work with the graphs we generated.

We decided we needed to create a standalone version of GyroSAT, so spreadsheet users could install it on their machine, to visualize the spreadsheets they work with. Therefore we adapted our implementation, and started to use the YFiles Graph Library for Windows Presentation Foundation for the graph visualization part. *Breviz* was born. We chose YFiles(available from yworks.com) since it supports hierarchical graphs out of the box, and provides great support for customizable user interaction.

Figure 2 shows a screen shot of the global view of a data flow diagram in Breviz.

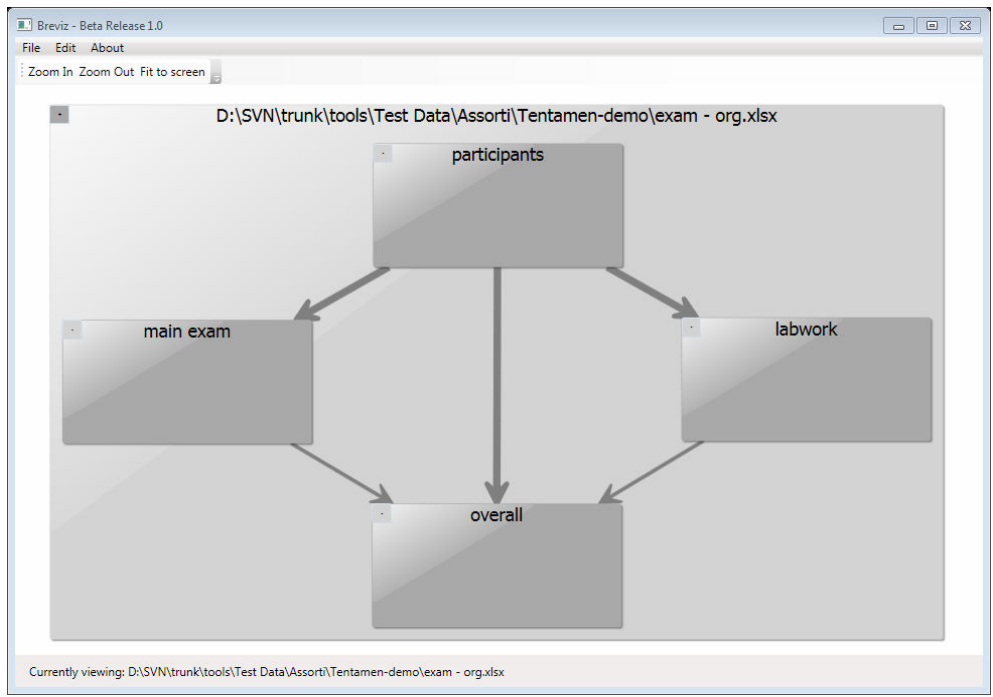

**Figure 2Breviz showing the global view of a simple spreadsheet with 4 worksheets**

## 5. EXAMPLES

While originally created to support spreadsheet users in transfer scenarios, we have seenthat Breviz is also useful for scenarios in which a spreadsheet user analyses the spreadsheet by itself. During an industrial case study at Dutch investment banker Robeco, we have found that the data flow diagrams really support users in understanding a spreadsheet. As one of the participants there stated ''this diagram (the global view was meant) shows me the idea behind the spreadsheet.'' Unfortunately we cannot describe the details of spreadsheets from the industrial case here, since they are confidential.

We therefore illustrate the usefulness of our approach by describing three cases that are loosely based on our experience in practice. Each of the three cases describes the application of one of the three views: global view, worksheet view and detailed view. The spreadsheets can be found on our website (http://swerl.tudelft.nl/bin/view/FelienneHermans/Publications)

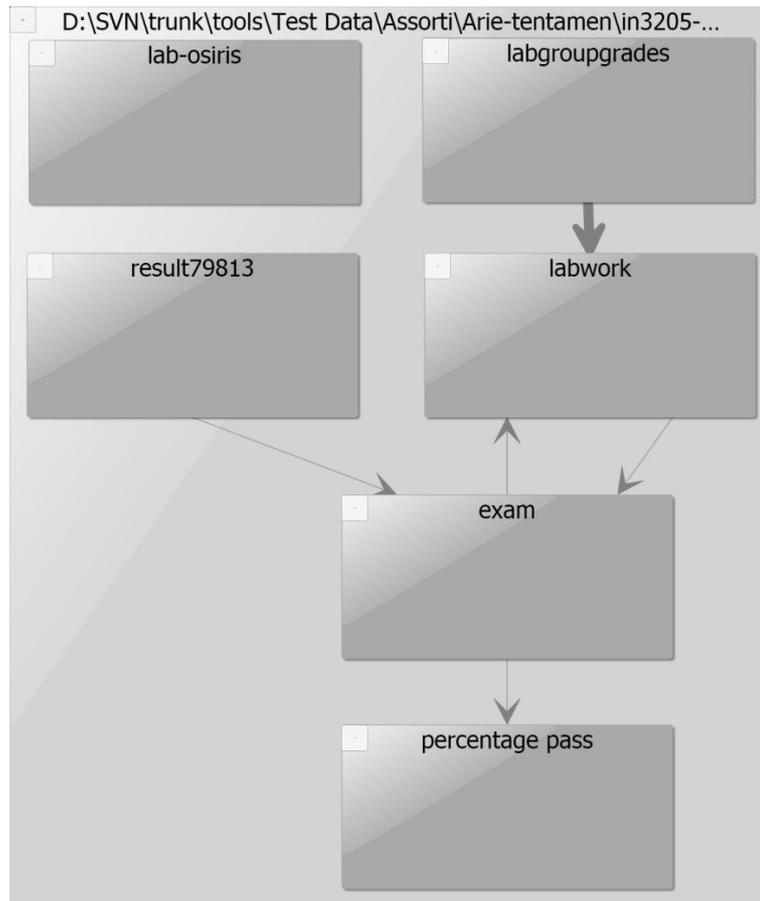

Figure 3 The global view for the Exam example

**5.1 Global view: Exam**
This spreadsheet is used by a university professor to calculate the grades for a course. It consists of six worksheets. From the global view, shown in Figure 3, some aspects of the spreadsheet immediately catch the eye. For instance, one of the worksheets 'lab-osiris' isnot connected to the other sheets, this sheet contains the data from Osiris, the university's grading system. This information can be crucial when working with the spreadsheet, since someone working with the spreadsheet might mistakenly think that all scores are updated when updating the information form Osiris. To determine this without Breviz would require the user to select all cells for all worksheets one by one and checking their dependents.
Furthermore the loop between 'exam' and 'labwork' stands out. The name 'exam' could suggest that the worksheet only contains data about the exam, however apparently also information regarding the lab work.

Besides helping to identify unexpected links between worksheets, the global view can serve as documentation for the spreadsheet. In the case study at Robeco we have seen that users paste an image of the global view in the spreadsheet to document it.

## 5.2 Worksheet view: Income statement

The second example is a worksheet that describes an income statement for a company.

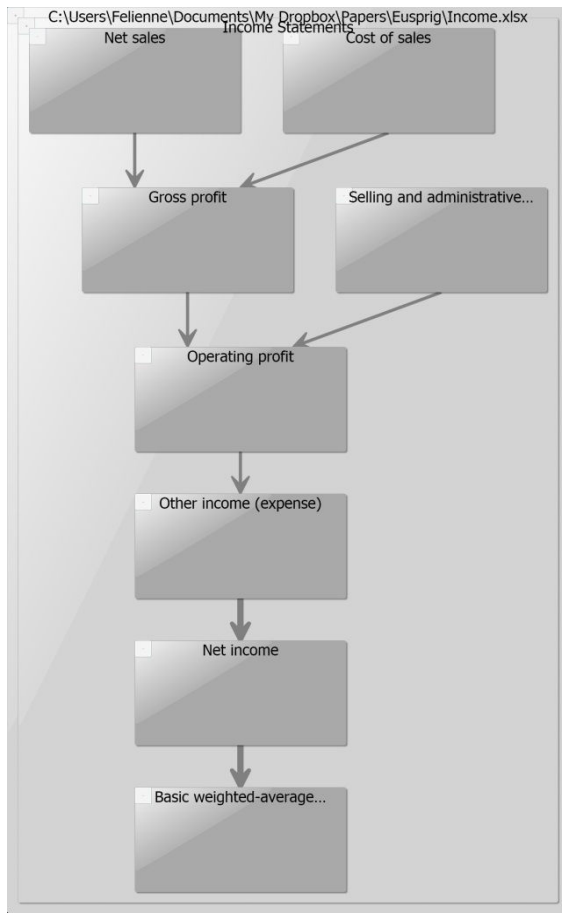

The worksheet is divided into a number of data blocks, but it is not immediately clear how they are related to each other.

Understanding the relation between the data blocks however contributes to understanding the worksheet as a whole. Figure 4 shows the worksheet view for this worksheet. From this diagram a spreadsheet user can immediately see how the data blocks are related. The diagram shows the ordering of the data blocks in the calculation. For instance, 'net sales' and 'cost of sales' are at the same level in the calculation, while in the worksheet one is located above the other, possibly confusing the spreadsheet user.

Since Breviz gathers names for the data blocks, the user can also see---to a certain extent--- whether the relations between make sense.

**Figure 4 Worksheet view for the Income example**

## 5.3 Formula view: Financial Performance

In the third example the financial performance of a company is calculated. This calculation is quite complicated, and to get an overview of what is calculated, spreadsheet users usually click all the cells to view the formulas. Breviz's formula view is meant to support the user in understanding the calculation easier. Figure 5 shows the formula view for this example.

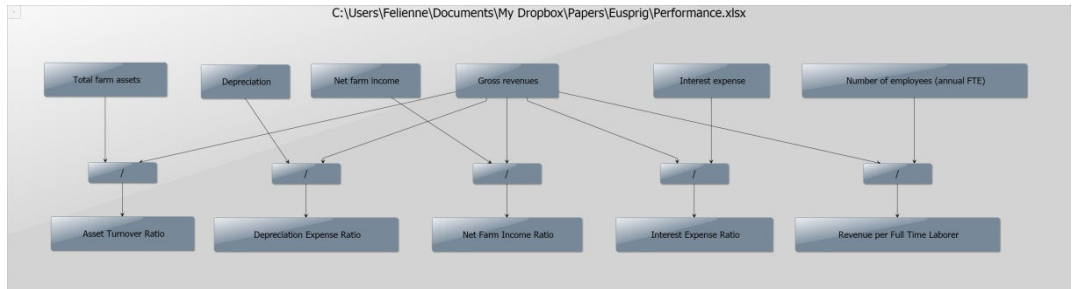

**Figure 5 Formula for the Performance example**

Without the need to inspect all formulas, spreadsheet users can see what exactly is calculated in the formulas. Breviz users where very satisfied with the help of the formula view. Many of them stated that it saves them a huge amount of time when analyzing a spreadsheet.

## 6. DISCUSSION

Section 5 gives an overview of the applicability of our approach for understanding and documenting spreadsheets. In this section we describe an additional application of Breviz we envision, but that has not been evaluated thoroughly in practice, namely the use of Breviz diagrams to detect anomalies in a spreadsheet.
As shown in Section 5.1 sometimes the data flow diagram gives rise to questions about the structure of the spreadsheet. This especially is the case in the global view, since we could see that level as the architecture of the spreadsheet, so it is logical to also analyze the quality of that architecture, as is common with software architecture.
Here we list a number of those *spreadsheet structure smells* of which we suspect that they might indicate errors. As stated before, this idea still needs more attention and a subsequent empirical evaluation.

- A loop between two worksheets: Data is going back and forth between two worksheets. This might indicate that the spreadsheet is not structured in a logical way, increasing the change of errors.

- A single arrow 'against the stream': All worksheets are connected is a single direction, but there is one link in the opposite direction. The one opposite dependency could be an error, or a special case that needs additional attention

- Worksheets that are disconnected from the data flow diagram. Users can mistakenly think the data in the disconnected sheet is taken into account in calculations in the other worksheets.

- Two worksheets that are very heavily coupled: the structure could possibly be improved by merging the worksheets. On a side note: merging heavily connected worksheets often has a positive impact on the performance of the spreadsheet, since worksheets with many connections tend to make the spreadsheet slow.

The further investigation of the effect of these spreadsheet smells is an interesting avenue for future research.

## 7. CONCLUSION

The goal of this paper is to explain how the technique of extracting a data flow diagram from a spreadsheet, which was created to support spreadsheet users in transfer scenarios, can be useful to help individual spreadsheet users to understand a spreadsheet quicker and better.

The current research gives rises to several directions for future work. Firstly the application of data flow diagrams for individual users could be refined. It might be the case that individual users have different information needs, so it would be logical to perform a new round of interviews in which we gather those information needs.

Furthermore the use of data flow diagrams to assess the quality of spreadsheets and spreadsheet design deserves more attention.